


\input vanilla.sty 


\mag=\magstep1
\hsize=6truein
\vsize=8.5truein
\TagsOnRight
\abovedisplayskip=9pt plus3pt minus6pt
\belowdisplayskip=9pt plus3pt minus6pt

\overfullrule=0pt

\font\tenbf=cmbx10
\font\tenrm=cmr10
\font\tenit=cmti10

\font\sevenbf=cmbx7
\font\sevenrm=cmr7
\font\sevenit=cmti7

\outer\def\heading{\bigbreak\bgroup\let\\=\cr\tabskip\centering
    \halign to \hsize\bgroup\bf\hfill\ignorespaces##\unskip\hfill\cr}
\def\endheading{\cr\egroup\egroup\nobreak\medskip}

\outer\def\endrmproclaim{\par\ifdim\lastskip<\medskipamount\removelastskip
  \penalty 55 \fi\medskip}

\def\beginref{
  \begingroup

  \textfont0=\sevenrm \scriptfont0=\fiverm \scriptscriptfont0=\fiverm
  \def\rm{\fam0 \sevenrm}
  \textfont1=\seveni  \scriptfont1=\fivei  \scriptscriptfont1=\fivei
  \def\mit{\fam1}
  \textfont2=\sevensy \scriptfont2=\fivesy \scriptscriptfont2=\fivesy
  \def\cal{\fam2}

  \def\it{\fam\itfam\sevenit} \textfont\itfam=\sevenit
  \def\bf{\fam\bffam\sevenbf} \textfont\bffam=\sevenbf
    \scriptfont\bffam=\sevenbf \scriptscriptfont\bffam=\fivebf
  \baselineskip=10pt

  \def\jour##1{{\it ##1\/}}
  \def\book##1{{\it ##1\/}}
  \def\vol##1{{\bf ##1}}

  \rm
}

\def\endref{\endgroup}

\def\=def{\; \mathop{=}_{\text{\rm def}} \;}
\def\del{\partial}
\def\res{\; \mathop{\text{\rm res}} \;}

\def\bR{{\bold R}}

\def\bP{{\bold P}}

\def\L{{\cal L}}
\def\M{{\cal M}}
\def\B{{\cal B}}
\def\T{{\cal T}}

\def\zbar{{\bar{z}}}

\def\Lhat{\hat{\L}}
\def\Mhat{\hat{\M}}
\def\Bhat{\hat{\B}}
\def\Fhat{\hat{F}}

\def\chat{\hat{c}}
\def\phat{\hat{p}}
\def\qhat{\hat{q}}
\def\uhat{\hat{u}}
\def\vhat{\hat{v}}
\def\what{\hat{w}}
\def\xhat{\hat{x}}
\def\yhat{\hat{y}}
\def\zhat{\hat{z}}

\pageno=1
\line{{\it College of Liberal Arts and Sciences} \hfill KUCP-0039}
\line{{\it Kyoto University} \hfill October 1991}
\vglue1truecm

\title
    Area-Preserving Diffeomorphisms and \\
    Nonlinear Integrable Systems
\endtitle
\author
    Kanehisa Takasaki \\
    {\rm Institute of Mathematics, Yoshida College, Kyoto University}\\
    {\it Yoshida-Nihonmatsu-cho, Sakyo-ku, Kyoto 606, Japan}\\
\endauthor
\vglue 2truecm

\noindent
{\bf Abstract\/}.
{\baselineskip=10pt
Present state of the study of nonlinear ``integrable" systems
related to the group of area-preserving diffeomorphisms
on various surfaces is overviewed.  Roles of area-preserving
diffeomorphisms in 4-d self-dual gravity are reviewed. Recent
progress in new members of this family, the SDiff(2) KP and Toda
hierarchies, is reported.  The group of area-preserving
diffeomorphisms on a cylinder plays a key role just as the infinite
matrix group GL($\infty$) does in the ordinary KP and Toda lattice
hierarchies. The notion of tau functions is also shown to persist
in these hierarchies, and gives rise to a central extension of
the corresponding Lie algebra. }


\vglue2truecm
\noindent
{\it
This article will be published in the proceedings of
the symposium ``Topological and geometrical methods in field theory,"
Turku, Finland, May 26 - June 1, 1991.
}

\newpage

\heading
    1. Introduction
\endheading

\noindent
The groups of area-preserving diffeomorphisms on various surfaces,
which we call SDiff(2) rather symbolically, give a natural extension
of the notion the group of circle diffeomorphisms Diff($S^1$).
This type of groups are known to emerge in a wide area of theoretical
and mathematical physics ranging
from fluid mechanics$^1$ 
and dynamical systems$^2$ 
to topics of high energy physics
such as $W_{\infty}$ algebras.$^3$ 

The notion of $W_\infty$ algebra is also intriguing from the standpoint
of the study of nonlinear integrable systems.  We already know that
self-dual gravity is characterized by an underlying SDiff(2) group
structure.   This should be by no means an isolated example;
looking for this type of nonlinear ``integrable" systems (the ``SDiff(2)
family," so to speak) is a challenging issue.  We expect to find some
other examples in a variety of models of field theories related
to $W_\infty$ algebras (or, more precisely, their``quasi-classical"
counterpart, i.e., $w_\infty$ algebras) such as:
$w_{\infty}$-gravity,$^4$ 
large-$N$ limit of nonlinear sigma models,$^5$ 
$N=2$ strings,$^6$ 
self-dual quantum gravity,$^7$ 
etc.

It is now widely recognized that various infinite dimensional Lie
algebras (and associated groups) play a key role in understanding
nonlinear integrables systems.$^8$ 
This fact has been well established for equations of KdV type;
associated Lie algebras are Kac-Moody algebras.  The KP hierarchy
as well as its Toda lattice version (Toda lattice hierarchy) are
known to be characterized by the gl($\infty$) algebra of
$\infty \times \infty$ matrices.  These equations are called
``soliton equations."

Similar structures have been observed for the 4-d self-dual
Yang-Mills equation and their dimensional reductions, the Bogomolny
equation (3-d), the principal chiral models and the Ernst equation.
These gauge field equations, too, are shown to have underlying
infinite dimensional algebras similar to Kac-Moody algebras.$^9$
(In fact, it is natural to enlarge these algebras by adding
derivation operators.$^{10,11}$) 
The situation is, however, somewhat different from soliton
equations; more than one variables are allowed to take place in its
loop algebra structure. In 4-d, there are indeed two extra variables
along with a ``spectral parameter" that also arises in soliton equations.
These variables may be thought of as local coordinates of
a three-dimensional complex manifold called ``twistor space."$^{12}$
In 3-d reductions, the twistor space is reduced to a two-dimensional
complex manifold called a ``minitwistor space"; this is the stage
where the Bogomolny equation is treated as a preliminary step
towards the principal chiral models and the Ernst equation.$^{13}$
In 2-d reductions, one will have a Riemann surface (mostly, a
Riemann sphere) that should be called a ``miniminitwistor space";
soliton equations of the KdV type are shown to fall into this class.$^{14}$

The situation further drastically changes in the case of self-dual gravity
(the vacuum Einstein equation). The role of algebras of Kac-Moody type
(with extra variables, if necessary) is now played by algebras of
diffeomorphisms.  According to the nonlinear graviton construction
of Penrose,$^{15}$ 
any self-dual vacuum Einstein space-time has a one-to-one
correspondence with a three dimensional complex manifold
(``curved twistor space").  This curved twistor space has
a projection (fibering) over a Riemann sphere, and each fiber
is given a symplectic structure that deforms as the point of
the Riemann sphere moves.  This is exactly the place where a group
of area-preserving diffeomorphisms emerges; those area-preserving
diffeomorphisms may also depend on a parameter $\lambda$ that takes
values in the Riemann sphere.  It is indeed shown that a loop algebra
of SDiff(2), like Kac-Moody-like algebras, gives rise to deformations
of curved twistor spaces, hence symmetries on the space of solutions of
self-dual gravity.$^{16,17}$
We shall first review these stories.

One may naturally ask if the pattern of dimensional reductions in gauge
field equations persists in this case.  In 3-d, this is indeed the case
as pointed out by Park.$^5$  Self-dual gravity has a 3-d reduction that
allows SDiff(2) on a cylinder to act as symmetries.  This significant
observation of Park has been extended to an SDiff(2) version of both
the KP hierarchy and the Toda lattice hierarchy, and yielded several
remarkable results. These results will be reported in the latter half
of this article. It seems likely that 2-d reductions of these equations
will fall into special (solvable) cases of the classical theory of 2-d
Monge-Amp\`ere equation, hence less exciting if compared with diverse
2-d reductions of the self-dual Yang-Mills equation. Nevertheless they
will still have rich contents in the context of $w_\infty$-gravity.$^{18}$

\heading
    2. Self-dual gravity and nonlinear graviton construction
\endheading

\noindent
We start with a brief review of the situation in 4-d self-dual gravity.
In 4-d, any vacuum Einstein metric has (locally) a complex K\"ahler
structure, and becomes a hyper-K\"ahler manifold.  In particular,
there are three independent K\"ahler structures and they can be mixed
by the unit quaternion group SU(2).$^{19}$  
This family of K\"ahler structures and associated K\"ahler forms
play a key role in the nonlinear graviton construction of Penrose.$^{15}$

A more down-to-earth formulation of this basic observation is presented
by Plebanski$^{20}$ 
and summarized in a review by Boyer.$^{21}$ 
(Essentially the same formulation is independently discovered by
Gindikin$^{22}$ 
from the standpoint of integral geometry.)
In the formulation of Plebanski (as well as of Penrose), one starts from
a (complexified) metric of the form
$$
    ds^2 = \det\pmatrix e^{11} & e^{12} \\
                        e^{21} & e^{22} \\
           \endpmatrix
         = e^{11}e^{22}-e^{12}e^{21}              \tag 1
$$
where $e^{ab}$ are linearly independent differential 1-forms.
This induces new local SL(2)$\times$SL(2) gauge symmetries that act
on both sides of the above $2 \times 2$ matrix of 1-forms.
With this gauge freedom, one can reduce the Ricci-flatness condition
to the closedness
$$
    d\omega^{cd} = 0                              \tag 2
$$
of the exterior 2-forms
$$
    \omega^{cd} = \omega^{dc}
    = \frac{1}{2}\in_{ab} e^{ac} \wedge e^{bd}    \tag 3
$$
where $\in_{ab}$ is the symplectic form normalized as
$$
    (\in_{ab}) = \pmatrix  0 & 1 \\
                          -1 & 0 \\
                 \endpmatrix                      \tag 4
$$
and the Einstein summation convention is applied to the symplectic
indices $a,b,\ldots$.

These equations can be cast into a single equation if one introduces
a complex parameter $\lambda$ and makes the linear combination
$$
    \omega(\lambda) = \omega^{11} +2\omega^{12}\lambda
                    + \omega^{22}\lambda^2.       \tag 5
$$
This is a realization of the aforementioned SU(2) mixing of different
K\"ahler forms, and $\lambda$ plays the role of a mixing parameter
ranging over a Riemann sphere (which is locally diffeomorphic to the
group manifold of SU(2)). Note that $\omega(\lambda)$ can also be written
$$
    \omega(\lambda) = \frac{1}{2} \in_{ab}
               (e^{a1}+e^{a2}\lambda) \wedge (e^{b1}+e^{b2}\lambda),
                                                  \tag 7
$$
hence
$$
    \omega(\lambda) \wedge \omega(\lambda) = 0.   \tag 8
$$
In terms of $\omega(\lambda)$, the three closedness equations can
be rewritten
$$
    d'\omega(\lambda) = 0,                        \tag 9
$$
where ``$d'$" means the total differentiation only with respect to
space-time coordinates; $\lambda$ is considered a constant as
$d'\lambda=0$. Eqs. 7 and 8 mean that there are a pair of
``Darboux coordinates" $P(\lambda)$ and $Q(\lambda)$ such that
$$
    \omega(\lambda) = d'P \wedge d'Q.             \tag 10
$$
Of course these Darboux coordinates are by no means unique; further,
they exist only locally, in particular with respect to $\lambda$.
As Penrose$^{15}$ first pointed out (in the form of an inverse
construction), these Darboux coordinates may be thought of as
a local expression of sections of the fibering
$$
    \pi: \ \T \to \bP^1                           \tag 11
$$
where $\T$ is the corresponding curved twistor space.

To see this correspondence more explicitly,$^{16,21}$ let us assume that
there are two sets of Darboux coordinates $(u(\lambda),v(\lambda))$ and
$(\uhat(\lambda),\vhat(\lambda))$ with the following Laurent
expansion in a neighborhood of a circle $|\lambda|=\rho$ on the complex
$\lambda$ plane.
$$
\align
      u(\lambda) = p\lambda +x + \sum_{n\le -1} u_n\lambda^n,
      \quad
 &    v(\lambda) = q\lambda +y + \sum_{n\le -1} v_n\lambda^n, \\
   \uhat(\lambda) = \phat + \xhat\lambda + \sum_{n\ge 2}\uhat_n\lambda^n,
   \quad
 & \vhat(\lambda) = \qhat + \yhat\lambda + \sum_{n\ge 2}\vhat_n\lambda^n.
                                                  \tag 12     \\
\endalign
$$
[The first two Laurent coefficients of these four functions are given
special status, because they are exactly the space-time coordinates
that arise in Plebanski's ``heavenly equations."$^{20}$
There are three different choices, $(p,q,\phat,\qhat)$, $(x,y,p,q)$ and
$(\xhat,\yhat,\phat,\qhat)$, that fit into the first heavenly equation,
the second heavenly equation, and the ``dual" second heavenly equation,
respectively.] The above situation is indeed the case illustrated by
Penrose.$^{15}$ Note that due to the special form of the Laurent series,
$u(\lambda)$ and $v(\lambda)$ can be analytically continued to the
outside of the circle and have first order poles at $\lambda=\infty$,
whereas $\uhat(\lambda)$ and $\vhat(\lambda)$ to the inside.  Meanwhile,
these two pairs of Darboux coordinates should be related by a canonical
(i.e., area-preserving) diffeomorphism as
$$
    f(\lambda,u(\lambda),v(\lambda)) = \uhat(\lambda), \quad
    g(\lambda,u(\lambda),v(\lambda)) = \vhat(\lambda),      \tag 13
$$
where $f$ and $g$ are holomorphic functions of three variables, say
$(\lambda,u,v)$, and satisfy the area-preserving condition
$$
    \frac{\del f(\lambda,u,v)}{\del u} \frac{\del g(\lambda,u,v)}{\del v}
   -\frac{\del f(\lambda,u,v)}{\del v} \frac{\del g(\lambda,u,v)}{\del u}
    = 1.                                                    \tag 14
$$
Geometrically, the pair of functions $f$ and $g$ (after appropriate
``twisting" by $\lambda$) give patching functions of local coordinates
on the twistor space $\T$, and the given Ricci-flat K\"ahler metric is
(at least locally) now encoded into this data.  Penrose$^{15}$ further
argues that this is a one-to-one correspondence, ensuring the converse
construction with the aid of the Kodaira-Spencer deformation theory
of complex manifolds. Analytically, this amounts to a kind of
``Riemann-Hilbert problem" now setup in the group of area-preserving
diffeomorphisms.$^{14,19}$ (Actually, here is a technical subtlety that
should be taken into account to establish a true one-to-one
correspondence; let us postpone it to the next section.)
The variable $\lambda$ now plays the role of a loop parameter,
hence the fundamental group structure is not of SDiff(2) but of
the loop group of SDiff(2) (on a plane).

\heading
    3. Origin of SDiff(2) symmetries
\endheading

\noindent
SDiff(2) symmetries originate in the group structure of the data
$(f,g)$ (due to composition of diffeomorphisms). More precisely,
one starts from left and right translations on the SDiff(2) group
manifold of the form
$$
    (f,g) \longrightarrow
            \exp\left( \epsilon \{\Fhat,\cdot\} \right)
            \circ (f,g) \circ
            \exp\left( -\epsilon \{F,\cdot\} \right)      \tag 15
$$
generated by the Hamiltonian vector fields
$$
      \{F,\cdot\} = \frac{\del F}{\del u}\frac{\del}{\del v}
                    -\frac{\del F}{\del v}\frac{\del}{\del u},
    \quad
    \{\Fhat,\cdot\} = \frac{\del \Fhat}{\del\uhat}\frac{\del}{\del\vhat}
                     -\frac{\del \Fhat}{\del\vhat}\frac{\del}{\del\uhat},
                                                           \tag 16
$$
where $F=F(\lambda,u,v)$ and $\Fhat=\Fhat(\lambda,\uhat,\vhat)$ are
functions of three variables and assumed to have the same analyticity
properties as $f$ and $g$.  This should give rise to a one-parameter
family of transformations for solutions of the vacuum Einstein equation
via the fundamental relation
$$
     \omega(\lambda) = d'u(\lambda) \wedge d'v(\lambda)
                     = d'\uhat(\lambda) \wedge d'\vhat(\lambda).
                                                        \tag 17
$$
It is now convenient (and even essential) to understand the self-dual
vacuum Einstein equation as an enlarged system$^{23}$ 
with auxiliary unknown
functions $u(\lambda)$, $v(\lambda)$, $\uhat(\lambda)$, $\vhat(\lambda)$
obeying the above equations.  SDiff(2) symmetries, in fact, act on this
system rather than the original self-dual vacuum Einstein equation.

An infinitesimal form of these SDiff(2) symmetries will be obtained
by calculating the transformations to the first order of $\epsilon$ as
$$
\align
     u \longrightarrow u + \epsilon \delta u + O(\epsilon^2), \quad
  &  v \longrightarrow v + \epsilon \delta v + O(\epsilon^2),   \\
   \uhat \longrightarrow \uhat + \epsilon \delta \uhat + O(\epsilon^2),\quad
  &\vhat \longrightarrow \vhat + \epsilon \delta \vhat + O(\epsilon^2).
                                                       \tag 18  \\
\endalign
$$
The coefficients of $\epsilon$ then defines a linear map
$\delta=\delta_{F,\Fhat}$ that represents an infinitesimal SDiff(2)
symmetries.  We now have to mention a technical remark announced in the
last section:  The transformations of the Darboux coordinates are
to be determined only after selecting a reference space-time coordinate
system. Such a choice of coordinates remains arbitrary in the nonlinear
graviton construction and gives residual ``gauge" freedom; we have to
fix it. Our prescription of this ``gauge fixing" is to select one of
the aforementioned Plebanski coordinate systems and to require that
the transformations of $(u,v,\uhat,\vhat)$ leave invariant these
coordinates, i.e.,
$$
    \delta_{F,\Fhat}z = 0 \quad
    \text{\rm for all}\ z \ \text{\rm in reference coordinate system}.
                                                      \tag 19
$$
If, for example, $(x,y,p,q)$ are adopted as a coordinate system,
we have$^{17}$:
$$
    \delta_{F,\Fhat}w(\lambda)
    = \left\{
         F(\lambda,u(\lambda),v(\lambda))_{\le -1}
         -\Fhat(\lambda,\uhat(\lambda),\vhat(\lambda))_{\le -1},
         w(\lambda)
      \right\}_{x,y}                                             \tag 20
$$
for $w(\lambda)=u(\lambda),v(\lambda)$ and
$$
    \delta_{F,\Fhat}\what(\lambda)
    = \left\{
         \Fhat(\lambda,\uhat(\lambda),\vhat(\lambda))_{\ge 0},
         -F(\lambda,u(\lambda),v(\lambda))_{\ge 0},
         \what(\lambda)
      \right\}_{x,y}                                             \tag 21
$$
for $\what(\lambda)=\uhat(\lambda),\vhat(\lambda)$,
where $(\quad)_{\ge 0}$ and $(\quad)_{\le -1}$ stand for the
projection operators acting on Laurent series of $\lambda$ as
$$
    (\sum a_n\lambda^n)_{\ge 0} = \sum_{n\ge 0} a_n\lambda^n, \quad
    (\sum a_n\lambda^n)_{\le -1}= \sum_{n\le -1}a_n\lambda^n,    \tag 22
$$
and $\{\quad,\quad\}_{x,y}$ a Poisson bracket in $(x,y)$,
$$
    \{ F,G \}_{x,y} = \frac{\del F}{\del x}\frac{\del G}{\del y}
                     -\frac{\del F}{\del y}\frac{\del G}{\del x}. \tag 23
$$
In general, these symmetries satisfy the commutation relations
$$
    \left[ \delta_{F_1,\Fhat_1}, \delta_{F_2,\Fhat_2} \right]
    = \delta_{ \{F_1,F_2\}, \{\Fhat_1,\Fhat_2\} },                \tag 24
$$
thereby respect the Lie algebraic structure of SDiff(2). It is
shown$^{17}$ that these symmetries can be further extended to the
Plebanski ``key functions," and shown to obey the same commutation
relations. This is far from obvious because these key functions are
obtained as ``potentials," hence determined only up to an integration
constant.

It should be noted that the above construction is still mathematically
ambiguous; we have taken a circle $|\lambda|=\rho$ in an ad hoc way,
and assumed a priori that two distinct Darboux coordinates should
live on each side of this circle.  This is obviously not very elegant.
A more elegant formulation will be achieved in terms of sheaf cohomology
as Park$^{10}$ pointed out.

\heading
    4. SDiff(2) KP hierarchy
\endheading

\noindent
The SDiff(2) version of the KP hierarchy that we now consider
is due to Krichever.$^{24}$  
Instead of a pseudo-differential operator $L$ in the KP hierarchy,
one starts from a Laurent series $\L$ of $\lambda$,
$$
    \L = \lambda + \sum_{i=1}^\infty u_{i+1}\lambda^{-i},       \tag 25
$$
where the Laurent coefficients are unknown functions of an infinite
number of variables $t=(t_1,t_2,\ldots)$, $t_1=x$. The first variable
$t_1$ is identified with a 1-d space variable  $x$ as in the
ordinary KP hierarchy. Krichever's hierarchy consists of the
evolution equations (``dispersionless Lax equations")
$$
  \frac{\del\L}{\del t_n} = \{ \B_n, \L \}_{\lambda,x},
  \quad n=1,2,\ldots,                                           \tag 26
$$
where $\B_n$ are polynomials of $\lambda$ given by
$$
    \B_n = (\L^n)_{\ge 0},                                      \tag 27
$$
$(\quad)_{\ge 0}$, as already mentioned, stands for the polynomial
part of a Laurent series of $\lambda$;
$\{\quad, \quad\}_{\lambda,x}$ now denotes a Poisson bracket with
respect to $(\lambda,x)$,
$$
    \{ F, G \}_{\lambda,x} = \frac{\del F}{\del\lambda}\frac{\del G}{\del x}
                -\frac{\del F}{\del x}\frac{\del G}{\del\lambda}.  \tag 28
$$
Obviously, this hierarchy is a kind of ``quasi-classical" version
to be obtained from the ordinary KP hierarchy by replacing
$$
\align
      \del/\del x  &\longrightarrow \lambda,                      \\
    [\quad, \quad] &\longrightarrow \{\quad, \quad\}_{\lambda,x}.
                                                          \tag 29 \\
\endalign
$$
The term ``SDiff(2)" now refers to the group of symplectic diffeomorphisms
that leave invariant the above Poisson bracket.  This is the group of
area-preserving diffeomorphisms on a cylinder on which $(\arg\lambda,x)$
give a coordinate system. Krichever seems to have been led to this
hierarchy from two routes: One is the route from the study of
``hydrodynamic" Hamiltonian structures and the``averaging method" in
soliton theory.$^{25}$
The other is the route from exact solutions of
``topological minimal models."$^{26}$
The above hierarchy contains the so called ``dispersionless KP equation"
in the 3-d sector $(x,y,t)=(t_1,t_2,t_3)$,$^{25}$ which is the same as the
ordinary KP (2-d KdV) equation except that a dispersion term is dropped.

Our approach to this new hierarchy$^{27}$ 
is more close to the work of
the Kyoto group.$^{28}$ 
A goal is to reorganize everything in terms of a ``tau function" and
an underlying Lie algebra. To this end, however, we have to borrow
crucial ideas from the method developed for the self-dual vacuum
Einstein equations.  We first look for a K\"ahler-like 2-form;
then introduce a pair of ``Darboux coordinates";
this leads to a twistor theoretical description (i.e., a nonlinear
graviton construction) of general solutions; infinitesimal variations
of ``patching functions" (which give a SDiff(2) group element)
give rise to infinitesimal symmetries of the hierarchy.  The use of
a pair of ``Darboux coordinates," one of which is the above $\L$
itself and the other is written $\M$ below, is a characteristic of
our approach. Our definition of the tau function is based upon these
two functions.

It deserves to be mentioned that the ordinary KP hierarchy, too, has
a counterpart of $\M$ that plays the role of a second Lax operator.
Such an improved Lax formalism of the KP hierarchy is introduced by
Orlov$^{29}$ 
and later applied to $d=1$ string theory by Awada and Sin.$^{30}$
This seems to suggest a possibility to interpret the well known relation
of the KP hierarchy and an infinite dimensional Grassmannian
manifold$^{28,31}$ 
as a kind of {\it noncommutative (mini)twistor theory}.

The rest of this section is devoted to a more detailed account of
our approach to the SDiff(2) KP hierarchy. In our definition,
the hierarchy consists of the Lax equations
$$
    \frac{\del\L}{\del t_n} = \{ \B_n, \L \}_{\lambda,x}, \quad
    \frac{\del\M}{\del t_n} = \{ \B_n, \M \}_{\lambda,x}     \tag 30
$$
and the canonical Poisson relation
$$
    \{ \L, \M \}_{\lambda,x} = 1,                            \tag 31
$$
where $\L$ is the same as explained above, and $\M$ is a Laurent series
(now expanded in powers of $\L$ rather than $\lambda$ for technical
reasons) of the form
$$
    \M = \sum_{n=1}^\infty nt_n\L^{n-1}
        +\sum_{i=1}^\infty v_{i+1}\L^{-i-1}.           \tag 32
$$
Thus there are two series of unknown functions, $u_i$ and $v_i$.
Note that we have excluded $u_1$ and $v_1$. In fact, one may include
these terms and consider the same equations; $u_1$ and $v_1$ then
turn out to be constants (independent of $t$), thereby can be absorbed
into redefinition of $\lambda$ and $\M$ as
$$
    \lambda  \to \lambda - u_1, \quad
    \M \to \M - v_1\L^{-1}.                            \tag 33
$$
One can therefore put $u_1=0$ and $v_1=0$ in the beginning. It is
however sometimes convenient to retain $v_1$ as a free parameter.

We then introduce a K\"ahler-like 2-form as
$$
    \omega = \sum_{n=1}^\infty d\B_n \wedge dt_n
           = d\lambda \wedge dx + \sum_{n=2}^\infty d\B_n \wedge dt_n,
                                                       \tag 34
$$
where ``$d$" now stands for total differentiation with respect to
{\it both} $t$ {\it and} $\lambda$. From the definition, $\omega$ is
a closed form,
$$
    d\omega = 0.                                       \tag 35
$$
Meanwhile, as in the case of the ordinary KP hierarchy,$^{28}$
the Lax equations for $\L$ turn out to be equivalent to the
``zero-curvature equations"
$$
    \frac{\del B_m}{\del t_n} - \frac{\del B_n}{\del t_m}
    + \{ B_m, B_n \}_{\lambda,x} = 0,                  \tag 36
$$
and these zero-curvature equations can be cast into the algebraic relation
$$
    \omega \wedge \omega = 0.                          \tag 37
$$
Eqs. 35 and 37 ensure the existence of two Darboux coordinates.
The fact is that $\L$ and $\M$ are such Darboux coordinates, i.e.,
they satisfy the fundamental equation
$$
    \omega = d\L \wedge d\M,                           \tag 38
$$
and, actually, this equation is an equivalent expression
of Eqs. 30 and 31.

Once we have arrived the above situation, it is very natural to
take another pair of Darboux coordinates $\Lhat$ and $\Mhat$ such that
$$
    \omega = d\Lhat \wedge d\Mhat                      \tag 39
$$
but with different Laurent expansion as
$$
    \Lhat = \sum_{n\ge 0}\uhat_i \lambda^i, \quad
    \Mhat = \sum_{n\ge 0}\vhat_i \Lhat^i.              \tag 40
$$
They should be linked with $\L$ and $\M$ by a pair of patching
functions $f=f(\lambda,x)$ and $g=g(\lambda,x)$
$$
    f(\L,\M) = \Lhat, \quad g(\L,\M) = \Mhat.          \tag 41
$$
The patching functions should satisfy the SDiff(2) condition
$$
    \{ f(\lambda,x),g(\lambda,x) \}_{\lambda,x} = 1.   \tag 42
$$
Actually, $\Lhat$ and $\Mhat$ are redundant variables and one may
characterize $f$ and $g$ as functions for which
$$
    f(\L,\M)_{\le -1} = 0, \quad g(\L,\M)_{\le -1} = 0. \tag 43
$$
Anyway, we thus have an SDiff(2) group element $(f,g)$, and
one can recover $(\L,\M)$ as a solution of a Rimann-Hilbert
problem as far as $(f,g)$ is sufficiently close to an identity
map. The action of a one-parameter group generated by a
Hamiltonian vector field
$$
    \{ F(\lambda,x), \cdot \}_{\lambda,x}
    = \frac{\del F(\lambda,x)}{\del\lambda}\frac{\del}{\del x}
     -\frac{\del F(\lambda,x)}{\del x}\frac{\del}{\del\lambda}
                                                        \tag 44
$$
gives rise to an infinitesimal symmetry $\delta_F$ of the SDiff(2)
KP hierarchy.  One can indeed obtain an explicit formula that resemble
the case of the self-dual vacuum Einstein equation:
$$
\align
    \delta_F\L &= \{ F(\L,\M)_{\le -1}, \L \}_{\lambda,x},           \\
    \delta_F\M &= \{ F(\L,\M)_{\le -1}, \M \}_{\lambda,x},   \tag 45 \\
\endalign
$$
and similar commutation relations:
$$
    \left[ \delta_{F_1}, \delta_{F_2} \right]
    = \delta_{ \{ F_1, F_2 \}_{\lambda,x} }.                 \tag 46
$$

Let us now turn to the problem of the tau function. In our approach,
the tau function is defined by the differential equations
$$
    \frac{\del\log\tau}{\del t_n} = v_{n+1},
    \quad n=1,2,\ldots,                                 \tag 47
$$
or, equivalently, by
$$
    d\log\tau = \sum_{n=1}^\infty v_{n+1}dt_n.          \tag 48
$$
The tau function therefore is always accompanied with an integration
constant:
$$
    \tau \longrightarrow c\tau, \ c \not= 0.            \tag 49
$$
Of course one has to prove that the right hand side of Eq. 48
is indeed a closed form.  This requires considerably technical
calculations exploiting the notion of formal residue,
$$
    \res \sum a_n \lambda^n d\lambda = a_{-1}.          \tag 50
$$
With the aid of such formal residue calculus, one can also prove that
the infinitesimal symmetries $\delta_F$ have a consistent extension
to the tau function. ``Consistency" means that extended symmetries
do not contradict the relation between $\log\tau$ and $v_{n+1}$.
Such an extension is given by
$$
    \delta_F\log\tau = -\res F^x(\L,\M) d_\lambda \L,   \tag 51
$$
where $F^x$ is a primitive function of $F=F(\lambda,x)$ with
respect to $x$ normalized as
$$
    F^x(\lambda,x) = \int_0^x F(\lambda,y)dy,           \tag 52
$$
and ``$d_\lambda$" stands for total differentiation with respect to
$\lambda$. A remakable consequence of this construction is that
the extended symmetries obey anomalous commutation relations:
$$
    \left[ \delta_{F_1}, \delta_{F_2} \right]\log\tau
    = \delta_{ \{F_1,F_2\}_{\lambda,x} }\log\tau + c(F_1,F_2),
                                                        \tag 53
$$
where
$$
    c(F_1,F_2) = \res F_1(\lambda,0)dF_2(\lambda,0).    \tag 54
$$
The anomalous term $c(F_1,F_2)$ gives a nontrivial cocycle of the
SDiff(2) algebra, hence a central extension.  This is reminiscent
of the case of the ordinary KP hierarchy; its tau function
has anomalous gl($\infty$) symmetries and leads to a central
extension of the infinite matrix algebra gl($\infty$).

Cocycles of SDiff(2) algebras on various surfaces are classified by
physicists.$^{32}$ 
According to their observations, there are $2g$ linearly independent
cocycles on a genus $g$ surface.  Since the present algebra lives
on a cylinder $S^1 \times \bR^1$, and this cylinder has genus $g=1/2$,
the above cocycle may be thought of as a realization of those predicted
cocycles.

\heading
    5. SDiff(2) Toda hierarchy
\endheading

\noindent
The ordinary 2-d Toda field theory on an infinite chain is described by
$$
    \del_z \del_\zbar \phi_i
    +\exp(\phi_{i+1}-\phi_i) -\exp(\phi_i-\phi_{i-1}) = 0,  \tag 55
$$
or, equivalently, for $\varphi_i=\phi_i-\phi_{i-1}$ by
$$
    \del_z \del_\zbar \varphi_i +\exp \varphi_{i+1}
    +\exp \varphi_{i-1} -2 \exp \varphi_i  = 0.             \tag 56
$$
In continuum limit as lattice spacing tends to 0, $\phi_i$ and
$\varphi_i$ will scale to 3-d fields $\phi=\phi(z,\zbar,s)$ and
$\varphi=\del\phi(z,\zbar,s)/\del s$, where $s$ is a coordinate
on the continuum limit of the infinite chain. Their equations of
motions are then given by
$$
    \del_z \del_\zbar \phi +\del_s \exp \del_s \phi = 0    \tag 57
$$
and
$$
    \del_z \del_\zbar \varphi +\del_s^2 \exp \varphi= 0.   \tag 58
$$
It is for this 3-d nonlinear field theory that Bakas$^3$ and Park$^5$
pointed out a w$_\infty$-algebraic structure. Saveliev and his
coworkers$^{33}$ 
attempted a different approach that exploits the notion of continual
Lie algebras, and presented a construction of solutions.  A Lax formalism
of the above equation, which contains a Poisson bracket, is proposed in
their work.  That type of Lax equations are also studied more systematically
by Golenisheva-Kutuzova and Reiman$^{34}$  
in the context of the coadjoint orbit method.

The above equations have two other sources.  One is discovered by
relativists.$^{35}$  
They pointed out that self-dual vacuum Einstein space-times
(``$\cal H$-spaces" or ``heavens" in their terminology) with a
rotational Killing symmetry are described by the above equation.
Another source is Einstein-Weyl geometry and associated curved
minitwistor spaces. This line is pursued in detail by twistor
people.$^{36}$  

We call the above equation (for $\phi$) the SDiff(2) Toda equation.
In this respect, the ordinary Toda equation may be called the
GL($\infty$) Toda equation.  The GL($\infty$) Toda equation
can be embedded into a Toda lattice version of the KP hierarchy,
i.e., the Toda lattice hierarchy.$^{37}$
Remarkably, the SDiff(2) Toda equation, too, has a similar
hierarchy, the ``SDiff(2) Toda hierarchy."$^{38}$
We now show a brief account of the theory of the SDiff(2) Toda
hierarchy.

We start from two pairs of Laurent series $(\L,\M)$ and $(\Lhat,\Mhat)$
of the form
$$
\align
     \L  = \lambda + \sum_{n\le 0} u_n\lambda^n, \quad
 &   \M  = \sum_{n\ge 1}nz_n\L^n + s + \sum_{n \le -1} v_n \L^n,  \\
   \Lhat = \sum_{n\ge 1} \uhat_n \lambda^n,      \quad
 & \Mhat = -\sum_{n\ge 1}n\zhat_n\Lhat^{-n}
                + s + \sum_{n \ge 1} \vhat_n\Lhat^n,         \tag 59  \\
\endalign
$$
where $z_n$ and $\zhat_n$, $n=1,2,\ldots$, now supply an infinite
number of independent variables along with $s$ and $\lambda$.
The hierarchy consists of the Lax equations
$$
    \frac{\del K}{\del z_n}     = \{ \B_n, K \}_{\lambda,s},  \quad
    \frac{\del K}{\del \zbar_n} = \{ \Bhat_n, K\}_{\lambda,s} \tag 60
$$
for $K=\L,\M,\Lhat,\Mhat$ and the canonical Poisson relations
$$
    \{ \L, \M \}_{\lambda,s} = \L, \quad
    \{ \Lhat, \Mhat \}_{\lambda,s} = \Lhat,                   \tag 61
$$
where $\B_n$ and $\Bhat_n$ are given by
$$
    \B_n    = (\L^n)_{\ge 0}, \quad
    \Bhat_n = (\Lhat^{-n})_{\le -1},                          \tag 62
$$
and the Poisson bracket is different from the SDiff(2) KP hierarchy:
$$
    \{ F, G \}_{\lambda,s}
    = \lambda\frac{\del F}{\del\lambda}\frac{\del G}{\del s}
     -\lambda\frac{\del G}{\del\lambda}\frac{\del F}{\del s}
                                                              \tag 63
$$
It is again crucial to introduce a K\"ahler-like 2-form as
$$
    \omega = \frac{d\lambda}{\lambda} \wedge ds
                +\sum_{n\ge 1} d\B_n \wedge dz_n
                +\sum_{n\ge 1} d\Bhat_n \wedge d\zhat_n.     \tag 64
$$
This satisfies the relations
$$
    d\omega = 0,  \quad  \omega \wedge \omega = 0,           \tag 65
$$
hence ensures the existence of Darboux coordinates. In fact,
$(\L, \M)$ and $(\Lhat, \Mhat)$ both give Darboux coodinates of
$\omega$,
$$
    \omega = \frac{d\L}{\L} \wedge d\M
           = \frac{d\Lhat}{\Lhat} \wedge d\Mhat,             \tag 66
$$
and this conversely characterizes the above defining equations of
the SDiff(2) KP hierarchy. The $\phi$-field can be reproduced
from the hierarchy as a potential:
$$
    d\phi = \sum_{n\ge 1}\res (\L^n d\log\lambda)dz_n
           -\sum_{n\ge 1}\res (\Lhat^{-n} d\log\lambda)d\zhat_n
           -\log \uhat_1 ds.                                 \tag 67
$$
(This resembles the Plebanski key functions.)  The tau function $\tau$
is also defined as a potential:
$$
    d\log\tau = \sum_{n\ge 1} v_{-n}dz_n
               -\sum_{n\ge 1} \vhat_n d\zhat_n + \phi ds.    \tag 68
$$

The nonlinear graviton construction now takes the following form.
The two pairs $(\L,\M)$ and $(\Lhat,\Mhat)$ of Darboux coordinates are
connected by patching functions $f=f(\lambda,s)$ and $g=g(\lambda,s)$
as
$$
    f(\L,\M) = \Lhat, \quad  g(\L,\M) = \Mhat,               \tag 69
$$
and the patching functions obey the SDiff(2) condition
$$
    \{ f(\lambda,s),g(\lambda,s) \}_{\lambda,s} = f(\lambda,s).
                                                             \tag 70
$$
The situation is thus almost parallel to the case of the self-dual
vacuum Einstein equation; one can obtain an infinitesimal symmetry
operator $\delta_{F,\Fhat}$ for a pair of generating functions
$F=F(\lambda,s)$ and $\Fhat=\Fhat(\lambda,s)$. Its action on
$\L$, $\M$, $\Lhat$, $\Mhat$ and $\phi$ can be calculated explicitly:
$$
\align
  \delta_{F,\Fhat}K
    &= \{ F(\L,\M)_{\le -1}-\Fhat(\Lhat,\Mhat)_{\le -1},
       K \}_{\lambda,s} \quad
    \text{for}\ K=\L,\M,                                            \\
  \delta_{F,\Fhat}K
    &= \{ \Fhat(\Lhat,\Mhat)_{\ge 0}
          -F(\L,\M)_{\ge 0},
       K \}_{\lambda,s} \quad
    \text{for}\ K=\Lhat,\Mhat.                                      \\
  \delta_{F,\Fhat}\phi
    &= -\res F(\L,\M)d\log\lambda +\res \Fhat(\Lhat,\Mhat)d\log\lambda.
                                                            \tag 71 \\
\endalign
$$
A consistent extension of these symmetries for the tau function can be
found:
$$
    \delta_{F,\Fhat}\log\tau
    = -\res F^s(\L,\M)d_\lambda\log\L
      +\res \Fhat^s(\Lhat,\Mhat)d_\lambda\log\Lhat,         \tag 72
$$
where
$$
    F^s(\lambda,s) = \int_0^s F(\lambda,t)dt, \quad
    \Fhat^s(\lambda,s) = \int_0^s \Fhat(\lambda,t)dt.
                                                            \tag 73
$$
The symmetries at the level of the tau function, again, obey
anomalous commutation relations as
$$
    \left[ \delta_{F_1,\Fhat_1}, \delta_{F_2,\Fhat_2} \right]\log\tau
    = \delta_{ \{F_1,F_2\}_{\lambda,s},
               \{\Fhat_1,\Fhat_2\}_{\lambda,s} }\log\tau
     +c(F_1,F_2) + \chat(\Fhat_1,\Fhat_2),                  \tag 74
$$
where $c$ and $\chat$ are cocycles of the SDiff(2) algebra given by
$$
\align
    & c(F_1,F_2) = -\res F_2(\lambda,0)dF_1(\lambda,0),         \\
    & \chat(\Fhat_1,\Fhat_2)
         = \res \Fhat_2(\lambda,0)d\Fhat_1(\lambda,0).  \tag 75 \\
\endalign
$$
This gives a central extension of the direct sum of two SDiff(2) algebras.

\heading
    6. Concluding remark
\endheading

\noindent
A problem left open is to find a geometric structure that should
correspond to the infinite dimensional Grassmannian manifold.$^{28,31,37}$
This should lead to a practical formula for the tau function
like the determinant formula and the field theoretical formula
already known for the KP and Toda lattice hierarchies.
Pursuing Orlov's approach$^{29,30}$ to the KP hierarchy is also an
intriguing problem. As mentioned in Section 4, Orlov's improved Lax
formalism may be thought of as a candidate of noncommutative
(mini)twistor theory. This might be related to geometric quantization
of $w_\infty$ gravity.

\heading
    Acknowledgements
\endheading
\noindent
I would like to express my gratitude to the organizing committee, in
particular, Professors Jouko Mickelsson and Osmo Pekonen for invitation
and hospitality.

\heading
    References
\endheading

\beginref

\item{1.}
Arnold, V., \jour{Ann. Inst. Fourier} \vol{16} (1966), 319-361.
\item{2.}
Moser, J., \jour{SIAM Review} \vol{28} (1986), 459-485.
\item{3.}
Bakas, I., \jour{Phys. Lett.} \vol{228B} (1989), 57-63;
\jour{Commun. Math. Phys.} \vol{134} (1990), 487-508.
\item{}
Pope, C.N., Romans, L.J., and Shen, X., \jour{Phys. Lett.} \vol{236B} (1990),
173-178; \jour{Phys. Lett.} \vol{245B} (1990), 72-78.
\item{4.}
Bergshoeff, E., Pope, C.N., Romans, L.J., Sezgin, E., Shen, X.,
and Stelle, K.S., \jour{Phys. Lett.} \vol{243B} (1990), 350-357.
\item{}
Bergshoeff, E., and Pope, C.N., \jour{Phys. Lett.} \vol{249B} (1990), 208-215.
\item{5.}
Park, Q-Han, \jour{Phys. Lett.} \vol{236B} (1990), 429-432;
\jour{Phys. Lett.} \vol{238B} (1990), 287-290.
\item{6.}
Ooguri, H., and Vafa, C., Self-duality and $N=2$ string magic,
Chicago preprint EFI-90-24 (April 1990).
\item{7.}
Yamagishi, K., and Chapline, F., \jour{Class. Quantum Grav.}
\vol{8} (1991), 1.
\item{8.}
\book{Vertex Operators and Physics},
Mathematical Sciences Research Institute Publications vol. 3;
\book{Infinite Dimensional Lie Groups with Applications}, ibid vol. 4
(Springer-Verlag, 1984).
\item{9.}
Chau, L.-L., Ge, M.-L., and Wu, Y.-S., \jour{Phys. Rev.} \vol{D25} (1982),
1086-1094; \jour{Phys. Rev.} \vol{D26} (1982), 3581-3592.
\item{}
Dolan, L., \jour{Phys. Lett.} \vol{113B} (1982), 387-390.
\item{}
Ueno, K., and Nakamura, Y., \jour{Phys. Lett.} \vol{109B} (1982), 273-278.
\item{}
Wu, Y.-S., \jour{Commun. Math. Phys.} \vol{90} (1983), 461-472.
\item{}
See also the following review and references cited therein:
\item{}
Chau, L.-L., in \book{Nonlinear Phenomena}, K.B. Wolf (ed.),
Lecture Notes in Physics vol. 189 (Splinger-Verlag  1983).
\item{10.}
Park, Q-Han, \jour{Phys. Lett.} \vol{257B} (1991), 105-110.
\item{11.}
Takasaki, K., \jour{Commun. Math. Phys.} \vol{127} (1990), 225-238.
\item{12.}
Ward, R.S., \jour{Phys. Lett.} \vol{61A} (1977), 81-82.
\item{}
Belavin, A.A. and Zakharov, V.E., \jour{Phys. Lett.} \vol{73B} (1978), 53-57.
\item{}
Chau, L.-L., Prasad, M.K. and Sinha, A., \jour{Phys. Rev.} \vol{D24} (1981),
1574-1580.
\item{13.}
Hitchin, N.J., \jour{Commun. Math. Phys.} \vol{89} (1983), 145-190.
\item{}
Ward, R.S., \jour{J. Math. Phys.} \vol{30} (1989), 2246-2251.
\item{}
Woodhouse, N.M.J., \jour{Class. Quantum Grav.} \vol{4} (1987), 799-814.
\item{14.}
Mason, L.J., and Sparling, G.A.J., \jour{Phys. Lett.} \vol{137A} (1989),
29-33.
\item{15.}
Penrose, R., \jour{Gen. Rel. Grav.} \vol{7} (1976), 31-52.
\item{16.}
Boyer, C.P., and Plebanski, J.F.,
\jour{J. Math. Phys.} \vol{26} (1985), 229-234.
\item{17.}
Takasaki, K., \jour{J. Math. Phys.} \vol{31} (1990), 1877-1888.
\item{18.}
Hull, C.M., The geometry of $W$-gravity,
Queen Mary and Westfield preprint QMW/PH/91/6 (June, 1991).
\item{19.}
Hitchin, N.J., Kahlhede, A., Lindstr\"om, U., and Ro\v{c}ek, M.,
\jour{Commun. Math. Phys.} \vol{108} (1987), 535-589.
\item{20.}
Plebanski, J.F., \jour{J. Math. Phys.} \vol{16} (1975), 2395-2402.
\item{21.}
Boyer, C.P., in \book{Nonlinear Phenomena}, K.B. Wolf (ed.),
Lecture Notes in Physics vol. 189 (Springer 1983).
\item{22.}
Gindikin, S.G., in \book{Twistor Geometry and Non-linear Systems},
H.D. Doebner and T. Weber (eds.), Lecture Notes in Mathematics
vol. 970 (Springer-Verlag 1982).
\item{23.}
Takasaki, K., \jour{J. Math. Phys.} \vol{30} (1989), 1515-1521.
\item{24.}
Krichever, I.M., The dispersionless Lax equations and topological
minimal models, preprint (April, 1991).
\item{25.}
Dubrovin, B.A., and Novikov, S.P., \jour{Soviet Math. Dokl.}
\vol{27} (1983), 665-669.
\item{}
Tsarev, S.P., \jour{Soviet Math. Dokl.} \vol{31} (1985), 488-491.
\item{}
Krichever, I.M., \jour{Funct. Anal. Appl.} \vol{22} (1989), 200-213.
\item{26.}
Dijkgraaf, R., Verlinde, H., and Verlinde, E., Topological strings in $d<1$,
Princeton preprint PUPT-1204, IASSNS-HEP-9o/71 (October 1990).
\item{}
Blok, B., and Varchenko, A., Topological conformal field theories and
flat coordinates, Princeton preprint IASSNS-HEP-91/05 (January 1990).
\item{27.}
Takasaki, K., and Takebe, T., SDiff(2) KP hierarchy, submitted to
\book{Proceedings of RIMS Research Project 91, Infinite Analysis},
RIMS, Kyoto University, June-August 1991.
\item{28.}
Sato, M., and Sato, Y.,  in \book{Nonlinear Partial Differential
Equations in Applied Sciences}, P.D. Lax, H. Fujita, and G. Strang (eds.)
(North-Holland, Amsterdam, and Kinokuniya, Tokyo, 1982).
\item{}
Date, E., Jimbo, M., Kashiwara, M., and Miwa, T.,
in \book{Nonlinear Integrable Systems}, M. Jimbo and T. Miwa (eds.)
(World Scientific, Singapore, 1983).
\item{29.}
Grinevich, P.G., and Orlov, A.Yu., in \book{Problems of modern quantum
field theory}, A. Belavin et al. (eds.) (Springer-Verlag, 1989).
\item{30.}
Awada, M., and Sin, S.J., Twisted $W_\infty$ symmetry of the KP hierarchy
and the string equation of $d=1$ matrix models, Florida preprint
UFITFT-HEP-90-33.
\item{31.}
Segal, G., and Wilson, G., \jour{Publ. IHES} \vol{61} (1985), 5-65.
\item{32.}
Arakelyan, T.A., and Savvidy, G.K.,
\jour{Phys. Lett.} \vol{214B} (1988), 350-356.
\item{}
Bars, I., Pope, C.N., and Sezgin, E.,
\jour{Phys. Lett.} \vol{210B} (1988), 85-91.
\item{}
Floratos, F.G., and Iliopoulos, J.,
\jour{Phys. Lett.} \vol{201B} (1988), 237-240.
\item{}
Hoppe, J.,
\jour{Phys. Lett.} \vol{215B} (1988), 706-710.
\item{33.}
Saveliev, M.V., and Vershik, A.M.,
\jour{Commun. Math. Phys.} \vol{126} (1989), 367-378.
\item{}
Kashaev, R.M., Saveliev, M.V., Savelieva, S.A., and Vershik, A.M.,
On nonlinear equations associated with Lie algebras of diffeomorphism
groups of two-dimensional manifolds,
Institute for High Energy Physics preprint 90-I (1990).
\item{34.}
Golenisheva-Kutuzova, M.I., and Reiman, A.G.,
\jour{Zap. Nauch. Semin. LOMI} \vol{169} (1988), 44 (in Russian).
\item{35.}
Boyer, C., and Finley, J.D., \jour{J. Math. Phys.} \vol{23} (1982),
1126-1128.
\item{}
Gegenberg, J.D., and Das, A., \jour{Gen. Rel. Grav.} \vol{16} (1984),
817-829.
\item{36.}
Hitchin, N.J., in \book{Twistor Geometry and Non-linear Systems},
H.D. Doebner and T. Weber (eds.), Lecture Notes in Mathematics  vol. 970
(Springer-Verlag 1982).
\item{}
Jones, P.E., and Tod, K.P., \jour{Class. Quantum Grav.} \vol{2} (1985),
565-577.
\item{}
Ward, R.S., \jour{Class. Quantum Grav.} \vol{7} (1990). L95-L98.
\item{}
LeBrun, C., Explicit self-dual metrics on $CP_2$ \# $\dots$ \# $CP_2$,
J. Diff. Geometry (to appear).
\item{37.}
Ueno, K., and Takasaki, K., in \book{Group Representations and Systems
of Differential Equations}, Advanced Studies in Pure Mathematics vol. 4
(Kinokuniya, Tokyo, 1984).
\item{}
Takebe, T., \jour{Commun. Math. Phys.} \vol{129} (1990), 281-318.
\item{38.}
Takasaki, K., and Takebe, T., SDiff(2) Toda equation
-- hierarchy, tau function and symmetries, Kyoto University
preprint RIMS-790 (August, 1991).

\endref
\bye